\documentstyle[12pt,sw20lart]{article}
\input{tcilatex}
\begin{document}

\author{Yi-shi Duan, Jun-ping Wang\thanks{%
Corresponding author; e-mail: wangjp0101@st.lzu.edu.cn}, Xin Liu
and
Peng-ming Zhang \\
%EndAName
{\small Institute of theoretical physics, Lanzhou University, Lanzhou
730000, P. R. China}}
\date{February 12, 2003}
\title{{\LARGE Topological Excitation in an Antiferromagnetic Bose-Einstein
Condensate}}
\maketitle

\begin{abstract}
\baselineskip0.5cm Two kinds of topological excitations, skyrmions
and monopoles, are studied. It is revealed that these two types of
excitations originate from different order parameter fields that
reflect different spatial distributions of the Neel vector
$\vec{m}$. Critical phenomena of the generation and annihilation
of skyrmion-antiskyrmion pairs are also discussed.
\end{abstract}

\section{Introduction}

The realization of a spinor Bose-Einstein condensate (BEc) \cite
{spinor,ss2,sss3} offords the opportunity to find richer topological
excitations in such dilute atomic gases. The spinor condensate is described
by a macroscopic wavefunction, $\Psi =\sqrt{n}\zeta ,$ where $n$ is the
total density of the gas, $\zeta $ is a normalized spinor that determines
the average local spin through the relation $\langle {\bf S}\rangle =\zeta
_a^{*}{\bf S}_{ab}\zeta _b$, and ${\bf S}$ represents the usual spin
matrices obeying the commutation relations $\left[ S_\alpha ,S_\beta \right]
=i\varepsilon _{\alpha \beta \gamma }S_\gamma \;$\cite{pra}. In the case of
spin-1 bosons, consideration of two ground states is necessary, because the
effective interaction between two spins can be either ferromagnetic or
antiferromagnetic \cite{machida,ho}. Also, there is a fundamental difference
between the possible topological excitations of the two states \cite{pra}.
In the antiferromagnetic case, the ground state energy is minimized for $%
\langle {\bf S}\rangle =0,$ and the spinor $\zeta $ is given by
\begin{equation}
\zeta =\frac{e^{i\vartheta }}{\sqrt{2}}\left(
\begin{array}{l}
-m^x+im^y \\
\;\;\;\;\sqrt{2}m^z \\
\;\;\;m^x+im^y
\end{array}
\right) ,  \label{spinor}
\end{equation}
where $\vartheta $ is the superfluid phase and $\vec{m}$ is the unit vector
field known as the N$\acute{e}$el vector \cite{stoof}. It can be seen from (%
\ref{spinor}) that the parameter space for the spinor $\zeta $ is $U\left(
1\right) \times S^2$, because both its overall phase and the spin
quantization axis can be chosen freely. Our interest is in the topological
excitations that are not analogous to the vortex excitations in the scalar
condensate and find $\pi _2\left( U\left( 1\right) \times S^2\right)
=e\oplus \pi _2\left( S^2\right) =Z.$ The meaning of this result is twofold.
It not only reveals the existence of a monopole excitation in a
three-dimensional antiferromagnetic BEc \cite{stoof} but also implies the
existence of a skyrmion excitation in the two-dimensional case \cite
{tsinghua,machida2}. Both types of excitations have been studied in other
fields of physics. Skyrmions appear in planar condensed matter systems that
can be described by the nonlinear sigma model (NSM) in the continuum limit,
including high temperature superconductors and systems exhibiting the
quantum Hall effect \cite{Hall}. Monopoles appear in gauge field theory \cite
{hooft}. However, the creation of such topological structures in BEc
presents an exciting opportunity to study the properties of skyrmions and
monopoles in exquisite detail, both theoretically and experimentally.

In this work, we study these two kinds of topological excitations, skyrmions
and monopoles, in an antiferromagnetic spin-1 BEc in the context of the $%
\phi $-mapping topological current theory \cite{Duan,ptp,D2,D3}. Because
these two kinds of excitations originate from different spatial
distributions of the N$\acute{e}$el vector field $\vec{m}$ , it is necessary
to introduce additional vector order parameter fields in order to study
these two excitations individually. It is shown that quantities such as
position, density and velocity of the two excitations expressed in terms of
the corresponding order parameter fields can be rigorously determined with
the $\phi $-mapping theory. Critical phenomena of the generation and
annihilation of skyrmion-antiskyrmion pairs are also discussed.

\section{Skyrmion excitation}

In this section we discuss the skyrmion excitation of a two-dimensional
antiferromagnetic BEc. The vector $\vec{m}$ sweeps the unit sphere $S^2$ an
integral number of times inside the core of a skyrmion and is uniform
outside the core. In analogy to previous work on skyrmions \cite{eg}, we
introduce a topological three-current as

\begin{equation}
J_s^\mu =\frac 1{8\pi }\varepsilon ^{\mu \nu \lambda }\varepsilon
_{abc}m^a\partial _\nu m^b\partial _\lambda m^c\;\;(\mu ,\nu ,\lambda
=0,1,2,\;\;a,b,c=1,2,3)  \label{sky}
\end{equation}
to describe the skyrmion excitations of the condensate, and its time
component is defined as the density of the total skyrmion charges, i.e. $%
J_s^0=\rho _s$. It is easy to prove that this current can be expressed as

\begin{equation}
J_s^\mu =\frac 1{8\pi }\varepsilon ^{\mu \nu \lambda }\left( \partial _\nu
A_\lambda -\partial _\lambda A_\nu \right) .  \label{wuyang}
\end{equation}
Here $A_\mu $ is the Wu-Yang potential \cite{WY}

\[
A_\mu =\vec{e}_1\cdot \partial _\mu \vec{e}_2,
\]
where $\vec{e}_1$ and $\vec{e}_2$ are two unit vectors normal to $\vec{m}$,
and $\left( \vec{e}_1,\vec{e}_2,\vec{m}\right) $ forms an orthogonal frame: $%
\vec{e}_1\times \vec{e}_2=\vec{m},$ $\vec{e}_1\cdot \vec{e}_2=0.$ Now,
consider a two-component vector field $\vec{\xi}=\left( \xi ^1,\xi ^2\right)
$ residing in the plane formed by $\vec{e}_1$ and $\vec{e}_2:$%
\[
e_1^i=\xi ^i/\left\| \xi \right\| ,\;\;\;e_2^i=\varepsilon _{ij}\xi
^j/\left\| \xi \right\|. \;\;\;( \left\| \xi \right\| =\sqrt{\xi ^j\xi ^j}%
,\;\;i,j=1,2)
\]
Then we have

\begin{equation}
A_\mu =\varepsilon _{ij}\frac{\xi ^i}{\left\| \xi \right\| }\partial _\mu
\frac{\xi ^j}{\left\| \xi \right\| }.\;\;\;  \label{wuyang2}
\end{equation}
With Eq. (\ref{wuyang2}), the current (\ref{wuyang}) is obtained as

\begin{equation}
J_s^\mu =\frac 1{4\pi }\varepsilon ^{\mu \nu \lambda }\varepsilon
_{ij}\partial _\nu \frac{\xi ^i}{\left\| \xi \right\| }\partial _\lambda
\frac{\xi ^j}{\left\| \xi \right\| }.  \label{curren t}
\end{equation}
Equation (\ref{curren t}) reveals the inherent conservation structure of the
skyrmion current:

\begin{equation}
\partial _\mu J_s^\mu =0.  \label{conserved}
\end{equation}
Within the $\phi $-mapping theory \cite{Duan,ptp}, the expression (\ref
{curren t}) is a topological current that can be rewritten in the compact
form

\begin{equation}
J_s^\mu =\frac 12\delta ^2\left( \vec{\xi}\right) D^\mu \left( \frac \xi
x\right) ,  \label{detlta}
\end{equation}
where \smallskip $D^\mu \left( \xi /x\right) $ is the vector Jacobian of $%
\vec{\xi},$

\begin{equation}
\varepsilon ^{ij}D^\mu \left( \frac \xi x\right) =\varepsilon ^{\mu \nu
\lambda }\partial _\nu \xi ^i\partial _\lambda \xi ^j,  \label{jacobian}
\end{equation}
and $D^0\left( \xi /x\right) $ is the usual two-dimensional Jacobian
determinant,

\[
D^0\left( \frac \xi x\right) =D\left( \frac \xi x\right) =\frac{\partial
\left( \xi ^1,\xi ^2\right) }{\partial \left( x^1,x^2\right) }.
\]
Equation (\ref{detlta}) is a new expression of the skyrmion three-current $%
J_s^\mu $, from which it can be seen that this current is non-vanishing only
at the zero points of $\vec{\xi}$:

\begin{equation}
\xi ^1\left( x^1,x^2,t\right) =0,\;\;\;\;\xi ^2\left( x^1,x^2,t\right) =0.
\label{eqs}
\end{equation}
Hence it is necessary to study the solutions of Eq. (\ref{eqs}) to determine
the nonzero solutions of $J_s^\mu $. If $D\left( \xi /x\right) =D^0\left(
\xi /x\right) \neq 0,$ the solutions of Eq. (\ref{eqs}) are

\begin{equation}
x^1=x_l^1\left( t\right) ,\;\;\;\;x^2=x_l^2\left( t\right)
,\;\;\;(l=1,2,\cdot \cdot \cdot ,N)  \label{solution}
\end{equation}
which \smallskip represent $N$ isolated zero points $\vec{z}_l\left(
t\right) \;(l=1,2,\cdot \cdot \cdot ,N)$ in space-time. These zero points
are the skyrmion excitations. The motion of the $l$th skyrmion is determined
by the $l$th world line $\vec{z}_l\left( t\right) .$

With the $\delta $-function theory \cite{delta}, it can be proved that

\[
\delta ^2\left( \vec{\xi}\right) =\sum_{l=1}^N\frac{\beta _l}{\left| D\left(
\xi /x\right) \right| _{_{\vec{z}_l}}}\delta ^2\left( \vec{r}-\vec{z}%
_l\left( t\right) \right) ,
\]
where the positive integer $\beta _l$ is called the Hopf index of the map $%
x\rightarrow \vec{\xi}.$ The meaning of $\beta _l$ is that when the point $%
\vec{r}$ covers the neighborhood of the zeros $\vec{z}_l$ once, the vector
field $\vec{\xi}$ covers the corresponding region $\beta _l$ times. Also,
with the definition of vector Jacobians (\ref{jacobian}) and using the
implicit function theorem \cite{D3}, the general velocity of the $l$th
skyrmion can be expressed as

\begin{equation}
V_l^\mu =\frac{dz_l^\mu }{dt}=\frac{D^\mu \left( \xi /x\right) }{D\left( \xi
/x\right) }\mid _{\vec{z}_l},\;\;\;V_l^0=1.  \label{skyvelodiyt}
\end{equation}
Then, the spatial and temporal components of \smallskip the skyrmion
current, $J_s^{1,2}$ and $J_s^0,$ can be respectively written in the forms
of the current and the density of a system of $N$ classical point particles
in (2+1)-dimensional space-time with topological charge $\frac 12W_l=\frac
12\beta _l\eta _l:$

\begin{eqnarray}
\vec{j}_s &=&\sum_{l=1}^N\frac{W_l}2\vec{V}_l\delta ^2\left( \vec{r}-\vec{z}%
_l\left( t\right) \right) ,  \label{skyfinal} \\
\rho _s &=&\frac 12\delta ^2\left( \vec{\xi}\right) D\left( \frac \xi
x\right) =\sum_{l=1}^N\frac{W_l}2\delta ^2\left( \vec{r}-\vec{z}_l\left(
t\right) \right) .  \nonumber
\end{eqnarray}
Here $W_l=\beta _l\eta _l$ is the winding number of the $\vec{\xi}$ field at
the zero point $\vec{z}_l\left( t\right) ,$ and $\eta _l=sgn(D\left( \xi
/x\right) \mid _{\vec{z}_l})=\pm 1$ is the Brouwer degree \cite{brouwer}: $%
\eta _l=+1\;$corresponds to skyrmion solutions, while $\eta _l=-1$
corresponds to antiskyrmion solutions. It is clear that Eq. (\ref{skyfinal})
describes the motion of the skyrmions in space-time. According to Eq. (\ref
{conserved}), the topological charges of these skyrmions are conserved:

\begin{equation}
\frac{\partial \rho _s}{\partial t}+\nabla \cdot \vec{j}_s=0.
\label{conserved2}
\end{equation}
The total charge of the system is

\[
\int \rho _sd^2x=\sum_{l=1}^N\frac{W_l}2.
\]
Note that, because the charge of a skyrmion is integer valued, the winding
number $W_l$ must take even number values \cite{78ho}.

\section{Generation and annihilation of skyrmion-antiskyrmion pairs}

The preceding analysis shows that the zeros of the vector field $\vec{\xi}$
play an important role in describing the skyrmions of an antiferromagnetic
BEc. The following analysis reveals a picture of great physical interest.
Note that the solutions (\ref{solution}) of Eq. (\ref{eqs}) are based on the
condition $D\left( \xi /x\right) =D^0\left( \xi /x\right) \neq 0.$ When this
condition does not hold, the result (\ref{solution}) will be altered \cite
{D2}. We denote one such zero point as $\left( t^{*},\vec{z}_l\right) ($%
i.e., $D\left( \xi /x\right) \mid _{\left( t^{*},\vec{z}_l\right) }=0$) and
assume that the Jacobian $D^1$ satisfies

\begin{equation}
D^1\left( \frac \xi x\right) \mid _{\left( t^{*},\vec{z}_l\right) }\neq 0.
\label{case1}
\end{equation}
From Eq. (\ref{skyvelodiyt}), it can be seen that at $\left( t^{*},\vec{z}%
_l\right) $, the velocity of the skyrmion is infinite:

\begin{equation}
\frac{dx^1}{dt}\mid _{\left( t^{*},\vec{z}_l\right) }=\frac{D^1\left( \xi
/x\right) }{D\left( \xi /x\right) }\mid _{\left( t^{*},\vec{z}_l\right)
}=\infty .  \label{limited}
\end{equation}
In order to explicitly analyze the behavior of the solutions of Eq. (\ref
{eqs}) near the point $\left( t^{*},\vec{z}_l\right) $, we can use the
Jacobian $D^1\left( \xi /x\right) $ instead of $D\left( \xi /x\right) $ for
the purpose of using the implicit function theorem. Then we have a unique
solution of Eq. (\ref{eqs}) in the neighborhood of $\left( t^{*},\vec{z}%
_l\right) ,$

\[
t=t\left( x^1\right) ,\;\;\;x^2=x^2\left( x^1\right) ,
\]
with $t^{*}=t\left( z_l^1\right) .$ It can be seen from (\ref{limited}) that

\[
\frac{dt}{dx^1}\mid _{\left( t^{*},\vec{z}_l\right) }=0.
\]
Then the Taylor expansion of $t=t\left( x^1\right) $ at the point $\left(
t^{*},\vec{z}_l\right) $ is

\begin{equation}
t-t^{*}=\frac 12\frac{d^2t}{dx_1^2}\mid _{\left( t^{*},\vec{z}_l\right)
}\left( x^1-z_l^1\right) ^2,  \label{para}
\end{equation}
which is a parabola in the $x^1$-$t$ plane. From Eq. (\ref{para}), we obtain
two solutions, $x_1^1\left( t\right) $ and $x_2^1\left( t\right) $ , which
give two branch solutions (world lines of the skyrmions) of Eq. (\ref{eqs}).
If $\frac{d^2t}{dx_1^2}\mid _{\left( t^{*},\vec{z}_l\right) }>0,$ we have a
branch solution for $t>t^{*};$ otherwise, we have branch solutions for $%
t<t^{*}.$ These two cases are related to the origin and the annihilation of
a skyrmion-antiskyrmion pair. From (\ref{limited}), we see that the velocity
of skyrmions is infinite when they are being annihilated or generated at
limit points. The restriction on the conservation of the skyrmion current (%
\ref{conserved2}) implies that the topological charges of these two
skyrmions must be opposite, i.e.,

\[
\frac 12\beta _{l_1}\eta _{l_1}=-\frac 12\beta _{l_2}\eta _{l_2},
\]
which shows that $\beta _{l_1}=\beta _{l_2}$ and $\eta _{l_1}=-\eta _{l_2.}$

\section{Monopole excitation}

There are monopole excitations in three-dimensional antiferromagnetic BEc.
The vector $\vec{m}$ sweeps $S^2$ an integral number of times when once
traversing a path around the sphere in which a monopole is contained.
According to the results of Ref. \cite{stoof} and our previous work \cite
{duanmo}, we can deduce the monopole four-current

\begin{equation}
J_m^\mu =\frac 1{8\pi }\varepsilon ^{\mu \nu \lambda \rho }\varepsilon
_{abc}\partial _\nu m^a\partial _\lambda m^b\partial _\rho m^c.\;\;\;(\mu
,\nu ,\lambda ,\rho =0,1,2,3)  \label{monocurrten}
\end{equation}
It is clear that the current (\ref{monocurrten}) is conserved, i.e. $%
\partial _\mu J_m^\mu =0.$ Using the $\phi $-mapping theory \cite{ptp}, it
can be proved that the current (\ref{monocurrten}) has the compact form

\begin{equation}
J_m^\mu =\delta ^3\left( \vec{\phi}\right) D^\mu \left( \frac \phi x\right) ,
\label{monodelta}
\end{equation}
where $\vec{\phi}$ is a three-component vector field defined as

\begin{equation}
\frac{\phi ^a}{\left\| \phi \right\| }=m^a,\;\;\;\left\| \phi \right\| =%
\sqrt{\phi ^a\phi ^a},\;\;\;\;(a=1,2,3)  \label{fielddefine}
\end{equation}
and $D^\mu \left( \frac \phi x\right) $ is the vector Jacobian of $\phi
\left( x\right) $:

\begin{equation}
\varepsilon ^{abc}D^\mu \left( \frac \phi x\right) =\varepsilon ^{\mu \nu
\lambda \rho }\partial _\nu \phi ^a\partial _\lambda \phi ^b\partial _\rho
\phi ^c.  \label{monojacobian}
\end{equation}
It can be seen from Eq. (\ref{monodelta}) that the monopole four-current $%
J_m^\mu $ is non-vanishing only at the zero points of $\vec{\phi}:$

\begin{equation}
\phi ^a\left( x^1,x^2,x^3,t\right) =0.\;\;\;\;\;(a=1,2,3)  \label{monoeqs}
\end{equation}
The solutions of Eq. (\ref{monoeqs}) can be generally expressed as

\[
x^1=x_i^1\left( t\right) ,\;x^2=x_i^2\left( t\right) ,\;x^3=x_i^3\left(
t\right) ,\;\;\;(i=1,2,\cdot \cdot \cdot ,K)
\]
which represent the world lines of $K$ isolated zero points $\vec{z}_i\left(
t\right) \;\left( i=1,2,\cdot \cdot \cdot ,K\right) .$ These zero points are
the monopole excitations. The motion of the $i$th monopole is determined by
the $i$th world line $\vec{z}_i\left( t\right) .$

The $\delta $-function theory \cite{delta} demonstrates the relation

\[
\delta ^3\left( \vec{\phi}\right) =\sum_{i=1}^K\frac{\beta _i}{\left|
D\left( \phi /x\right) \right| _{_{\vec{z}_i}}}\delta ^3\left( \vec{r}-\vec{z%
}_i\left( t\right) \right) ,
\]
where \smallskip $\beta _i$ is the Hopf index of the map $x\rightarrow \vec{%
\phi}.$ With the definition of the vector Jacobians (\ref{monojacobian}),
and using the implicit function theorem, we can obtain the general velocity
of the $i$th monopole:

\[
V_i^\mu =\frac{dz_i^\mu }{dt}=\frac{D^\mu \left( \phi /x\right) }{D\left(
\phi /x\right) }\mid _{\vec{z}_i},\;\;\;V_i^0=1.
\]
Then, the monopole current $J_m^\mu $ can be written in the form of the
current and the density of a system of $K$ classical point particles in
(3+1)-dimensional space-time with topological charge $W_i=$ $\beta _i\eta _i$%
:

\begin{eqnarray}
\vec{j}_m &=&\sum_{i=1}^KW_i\vec{V}_i\delta ^3\left( \vec{r}-\vec{z}_i\left(
t\right) \right) ,  \label{monofinal} \\
\rho _m &=&\delta ^3\left( \vec{\phi}\right) D\left( \frac \phi x\right)
=\sum_{i=1}^KW_i\delta ^3\left( \vec{r}-\vec{z}_i\left( t\right) \right) ,
\nonumber
\end{eqnarray}
where $\eta _i=sgn(D\left( \phi /x\right) \mid _{\vec{z}_i})=\pm 1$ is the
Brouwer degree, and $W_i=\beta _i\eta _i$ is the winding number of $\vec{\phi%
}$ at the zero point $\vec{z}_i\left( t\right) .$ It is clear that Eq. (\ref
{monofinal}) describes the motion of the monopoles in space-time. Here, $%
\eta _i=+1$ corresponds to a monopole and $\eta _i=-1$ corresponds to an
antimonopole.

\section{Conclusion and discussion}

In conclusion, two kinds of topological excitations, skyrmions and
monopoles, in an antiferromagnetic BEc were studied in the context of the $%
\phi $-mapping theory. Because these two kinds of excitations originate from
different spatial distributions of the N$\acute{e}$el vector field $\vec{m},$
two vector order parameter fields, $\vec{\xi}$ and $\vec{\phi},$ defined in
terms of $\vec{m}$ were introduced in order to study these two excitations
separately. We found that these two kinds of excitations are generated from
the zero points of the corresponding vector order parameter fields and that
their topological charges are both characterized by the Hopf index and the
Brouwer degree. We also found that quantities such as the density and
velocity of these excitations can be rigorously determined using the $\phi $%
-mapping theory. Further investigation of the topology of the two-component
vector order parameter field $\vec{\xi}$ revealed physical pictures of the
generation and annihilation of skyrmion-antiskyrmion pairs. It was shown
that the velocity of the skyrmion is infinite in such critical processes.

Unlike previous works, we concentrated mainly on topological properties
rather than dynamical properties of skyrmions and monopoles. It was found
the analysis of the topology of the corresponding vector order parameter
fields of these excitations is sufficient to determine the topological
characteristics as well as their kinematics. In addition, we have elucidated
the inherent conservation structure of the skyrmion three-current and
studied the critical phenomena of the generation and annihilation of
skyrmion-antiskyrmion pairs on the basis of this structure. These results
have general significance for subsequent investigation of skyrmions.

To study the stability of these two types of excitations, the fact that
should be considered first is that they are different kinds of topological
structures: monopoles originate from natural singularities of the N$\acute{e}
$el vector field $\vec{m}$ in three-dimensional space, while skyrmions are $%
nonsingular$ excitations that originate from the nontrivial homotopy classes
of mappings from the compactified $R^2\sim S^2\rightarrow S^2$ in
two-dimensional space. Due to the singular nature of the spin texture of the
monopole, the condensate density vanishes in the core, and the monopole
turns out to be thermodynamically stable \cite{stoof}. This is analogous to
the case of vortex excitation in a scalar BEc. Contrastively, topology
allows the spin texture of a skyrmion to be of $arbitrary$ intrinsic size.
As a result, the stability of a skyrmion is determined by energetic
properties, not by topological properties \cite{tsinghua}.

\section*{Acknowledgements}

This work was supported by the National Natural Science Foundation and the
Doctor Education Fund of the Educational Department of China.

\end{document}